\documentclass[a4paper,11pt]{article}
\pdfoutput=1 % if your are submitting a pdflatex (i.e. if you have
\usepackage{jinstpub} % for details on the use of the package, please
\usepackage{lineno}
\title{200 mm Sensor Development Using Bonded Wafers}
\author[b]{M. Alyari}
\author[a]{R. Bradford}
\author[b]{M. Campanella}
\author[b]{P. Camporeale}
\author[e]{R. Demina}
\author[b]{J. Everts}
\author[b]{Z. Gecse}
\author[b]{R. Halenza}
\author[g]{U. Heintz}
\author[c]{S. Holland}
\author[f]{S. Hong}
\author[e]{S. Korjenevski}
\author[b]{A. Lampis}
\author[b]{R. Lipton}
\author[f]{R. Patti}
\author[d]{J. Segal}
\author[a]{K.W. Shin}

\affiliation[a]{Argonne National Laboratory, Lemont, Illinois 60439}
\affiliation[b]{Fermilab, P.O. Box 500, Batavia, Illinois USA}
\affiliation[c]{Lawrence Berkeley National Laboratory (LBNL),Berkeley, CA USA}
\affiliation[d]{SLAC National Accelerator Laboratory, Menlo Park, Ca USA}
\affiliation[e]{University of Rochester, Rochester, NY USA}
\affiliation[f]{NHanced Semiconductors, Naperville, IL USA}
\affiliation[g]{Brown University, Providence, RI USA}

\emailAdd{lipton@fnal.gov}

\title{200 mm Sensor Development Using Bonded Wafers}

\abstract {Sensors fabricated from high resistivity, float zone, silicon material have been the basis of vertex detectors and trackers for the last 30 years.  The areas of these devices have increased from a few square cm to $\> 200\ m^2$ for the existing CMS tracker.  High Luminosity Large Hadron Collider (HL-LHC), CMS and ATLAS tracker upgrades will each require more than $200\ m^2$ of silicon and the CMS High Granularity Calorimeter (HGCAL) will require more than $600\ m^2$. The cost and complexity of assembly of these devices is related to the area of each module, which in turn is set by the size of the silicon sensors. In addition to large area, the devices must be radiation hard, which requires the use of sensors  thinned to 200 microns or less.  The combination of wafer thinning and large wafer diameter is a significant technical challenge, and is the subject of this work. We describe work on development of thin sensors on $200 mm$ wafers using wafer bonding technology. Results of development runs with float zone, Silicon-on-Insulator and Silicon-Silicon bonded wafer technologies are reported.}

\begin{document}
%\setpagewiselinenumbers
%\linenumbers
\maketitle
\flushbottom
%\linenumbers

\section{Introduction}
Particle Physics collider experiments are increasingly dependent on silicon diode detectors for tracking, vertexing, 
and calorimetry.  These detectors can provide precise position and energy information and are sufficiently 
radiation hard for the challenging environment of the High Luminosity LHC.  ATLAS\cite{Collaboration:2017mtb} and CMS\cite{Collaboration:2272264}
 trackers at HL-LHC plan silicon tracking systems of $\approx 200 m^2$ each. The CMS high granularity calorimeter\cite{Collaboration:2293646} plans 
a detector of $600 m^2$ area tiled with planes of sensors diced from 8" wafers. In addition silicon-based sensors are increasingly 
utilized for x-ray imaging and other applications that will require large areas of sensors. Modern technologies 
such as 3-D integration often can only be affordably implemented on larger wafers. These needs motivate
the development of technologies to move sensor wafer fabrication from 6" to 8" wafers~\cite{Bergauer:2019jco}.

Sensors for high energy physics must have thin (100-300 micron) active regions for radiation resistance  
and low mass.  Typical 200 mm wafer processing equipment requires thicker (775 micron) material for automated handling and 
to limit breakage. These thick wafers must be thinned after topside processing to the 100-300 microns needed for HEP detectors. Good sensor performance with low leakage current requires a high quality backside contact.  This 
requires a p+ or n+ implant and associated annealing,  metalization, and sintering. The standard high temperature annealing process precludes the presence of topside metalization before the anneal.  There are several possible 
process solutions to this:
\begin{itemize}
\item The topside can be completely patterned and oven-annealed, topside metalization deposited. The wafer is then thinned, implanted and 
annealed using a laser that locally melts the backside implanted region\cite{Lipton:2009ug}.
\item The wafer can be implanted on front side, thinned, implanted on the back side, annealed, and metal deposited and 
patterned on the front side.  This involves handling and patterning a thin wafer during the final steps.
\item Use of alternative low temperature annealing processes such as microwave annealing\cite{YL}\cite{JS}.
\item The sensor (float zone) wafer backside can be implanted, polished, then Silicon-on-Insulator (SOI) bonded to a handle wafer for 
processing.  The top (device) wafer is then thinned to the required thickness and polished. After the topside process is complete the backside handle and oxide are removed and an aluminum backside contact electrode is deposited.
\item The sensor wafer can be Silicon-Silicon (SiSi) bonded to a low resistivity handle wafer of the same type.  The topside can be 
thinned to the appropriate thickness, polished and fully processed. The resulting stack can then be thinned to the required 
physical thickness.  No backside processing other than metalization is necessary.
\end{itemize}
We have chosen to explore the last two technologies as part of a US Small Business Innovative Research (SBIR)-funded development project aimed at production 
of large area, thinned, radiation hard silicon detector systems. The initial runs used the Novati foundry in Austin, Texas and 
the final run used the NHanced foundry in Morrisville, North Carolina.

\section{Detector Requirements}
This development was guided by the need for radiation hard detectors for Particle Physics.  Designs are driven by radiation 
induced effects, including the increase of acceptor levels, which require an increase in operating voltage, and the 
increase in leakage current\cite{Moll:1999kv}. These effects can be mitigated by making the detectors thin and operating them at low temperature.
The detectors planned for HL-LHC trackers are based on n-on-p diodes, thinned to a volume compatible with acceptable signal to noise after 
irradiation~\cite{TrackerGroupoftheCMS:2020zbt}.  At the  end-of-life the devices also must withstand bias voltages up to 1000 volts. The CMS HGCAL application required only DC-coupled 
pad sensors.  Tracker sensors incorporate AC coupling capacitors and polysilicon bias resistors on the microstrip sensors. 

\begin{table}[h]
\begin{tabular}{|l|l|l|l|l|l|}
\hline
Run & Substrate & \begin{tabular}[c]{@{}l@{}}Active/Physical \\ Thickness ($\mu m$)\end{tabular} & Process & Splits       & Date    \\ \hline
1   & FZ        & 725,500/725,500   & DC      & p-stop, oxide & 4/2015  \\ \hline
2   & SOI, FZ   & 200,500/700,500   & DC      & p-stop, oxide  & 3/2016  \\ \hline
3   & SOI       & 275/275       & AC      & p-stop,n+     & 11/2017 \\ \hline
4   & SiSi      & 200/700       & DC      & p-stop        & 3/2019  \\ \hline
\end{tabular}
\caption{Summary of process runs, substrates used and process splits.  Oxide splits used either wet or dry oxydation. P-stop splits varied 
the p-stop doping }
\label{tab:Runs}
\end{table}

\section{Process Development} 
The fabrication process is based on processes developed at SLAC and LBNL\cite{Holland:1988es}. These processes were adapted for the foundry 
process at NHanced. The process steps, including implantation and annealing were modeled using Silvaco tools\cite{Silvaco} at Fermilab. In total 
there were four runs, summarized in Table \ref{tab:Runs}. The first run used bulk float zone wafers to establish the overall process. Runs 2 and 3 used SOI stacked wafers and the last used SiSi wafer stacks. 
Initial development runs were DC-coupled diodes processed 
using full thickness wafers.  For the float zone wafers the minimum final thickness was 500 microns due to the fragility of thinner wafers. 
The initial runs were used to understand leakage current and breakdown characteristics, explore guard ring variants, 
and explore process splits such as p-stop dose and wet or dry oxidation. These also helped to inform the development of 
design rules for subsequent fabrications.  
The third run added AC coupling and polysilicon resistors. 

\begin{figure}[h]
\centering
\includegraphics[width=100 mm]{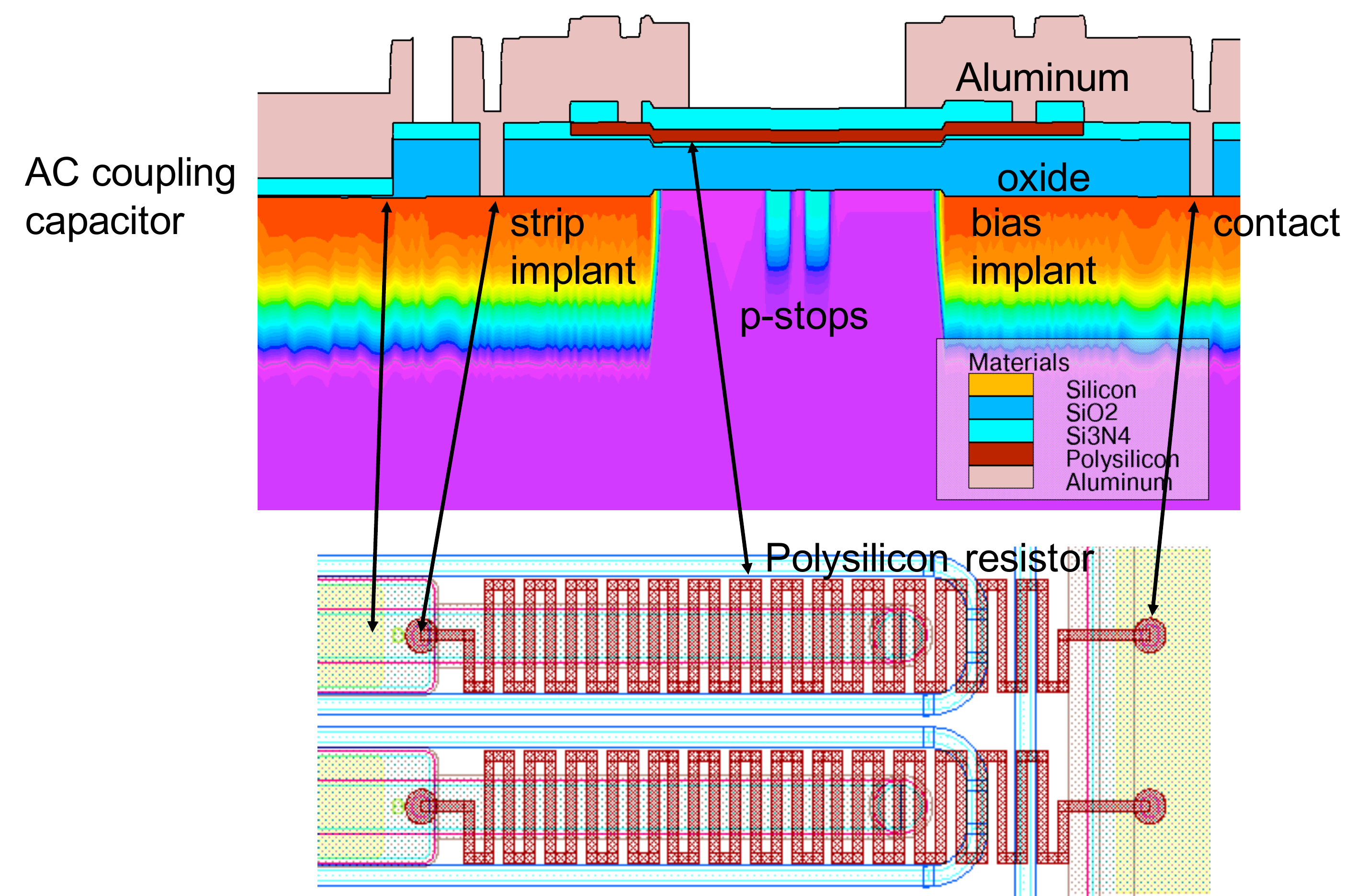}
\caption{Results of a simulation of the full AC wafer fabrication process with the corresponding structures on a typical 
silicon strip detector.}
\label{process}
\end{figure}

The overall process flow for DC and AC coupled variants is given in table \ref{tab:PFlow}. An initial sacrificial oxide is grown 
to getter out impurities.  Combined p-spray and p-stop isolation was used in the first run, and p-stops in the other runs. Another sacrificial oxide layer 
is used to provide the mask for all of the topside implants, providing good relative registration. The specific p, p+, or n
implants are then defined by photoresist.  The AC process includes steps to define the polysilicon resistors and coupling 
capacitors. The relative alignment of the mask layers is better than 3 microns. Figure~\ref{process} 
shows the result of a process simulation of the full AC/polysilicon process.

\begin{table}[h]
\begin{tabular}{|l|c|c|l|}
\hline
Process steps & DC  &  AC  & Comments    \\ \hline
Initial SOI, SiSi wafer preparation  & \textbullet  & \textbullet   & Implant (SOI), bond to handle  \\ \hline
Grow/Remove gettering oxide   & \textbullet  & \textbullet   &  Getter impurities \\ \hline
Blanket p-spray implant   & \textbullet &  & Not used for runs 3,4  \\ \hline
Grow masking oxide   & \textbullet     & \textbullet   &   \\ \hline
Pattern n and p implants in $SiO_2$  &  \textbullet   &  \textbullet  &  Oxide openings define n+ and p   \\ \hline
Pattern and implant  top &   \textbullet   &  \textbullet  & n+, p-stop, p-edge  \\ \hline
anneal  &  \textbullet    & \textbullet   & Remove implant oxide \\ \hline
Grow final oxide   &   \textbullet   &  \textbullet  &   \\ \hline
Dep/pattern/etch polysilicon & & \textbullet &   \\ \hline
Pattern/etch Capacitor Oxide & & \textbullet & Remove oxide for AC coupling\\ \hline
Dep/pattern/etch capacitor & & \textbullet &  $SiO_2-SiN-SiO_2$ dielectric \\ \hline
Pattern/etch contacts & & \textbullet &  Bias and resistor contacts \\ \hline
Dep/pattern/etch aluminum& \textbullet & \textbullet &  Top metal \\ \hline
Dep/pattern/etch passivation& \textbullet & \textbullet &  Top $SiO_2$ \\ \hline
Bond to top handle &  & \textbullet &  Thinned SOI process \\ \hline
Remove back handle, etch BOX&  & \textbullet &  Thinned SOI process \\ \hline
Deposit backside \emph{Al}, remove handle&  & \textbullet &  Thinned SOI process \\ \hline
\end{tabular}
\caption{Summary of process flows for DC and AC runs }
\label{tab:PFlow}
\end{table}

The SOI runs were fabricated with a 200 or 250 micron thick float zone device layer bonded to a 500 micron thick 
handle layer with $\approx 1$ micron of oxide separating the two layers.  A thick float zone wafer is backside implanted and then 
bonded to the handle.  The wafer stack is annealed at 1200 deg. C.  This anneal is also used to getter the sensor. After 
bonding the top wafer is polished down to the final active thickness. The wafer stack front side is then processed normally.
Run 2 used a DC-coupled process and wafers were delivered with the handle attached. 
These parts were biased from the topside contacts. Topside contacts are acceptable for applications 
with moderate radiation requirements.  At high radiation exposures the resistivity of the topside contact increases and may make topside bias connections problematic\cite{BaselgaBacardit:2316596}.  In run 3 a topside handle was attached, 
the backside handle wafer was removed, and the SOI buried oxide (BOX) was etched away.
The backside was then metalized.  In both cases standard float zone wafers were processed in parallel as control samples.

The last run used a Silicon-Silicon (SiSi) bonded substrate.  This consists of a high resistivity wafer directly bonded to a low 
resistivity handle such that the interface is transparent to charge carriers.  This construction has the advantage 
that the ohmic backside contact is built-in as part of the bonded wafer interface.  It eliminates the pre-bonding 
backside implant step and the post process backside handle removal and BOX etch steps needed for the SOI 
devices. The wafer simply needs to be ground down to the desired physical thickness and aluminized. Silicon-Silicon bonding 
is, however, a less well-established and controlled process.

Full investigation of the SiSi process was interrupted by the sale of the Novati foundry where Runs 1-3 were fabricated. This required 
re-qualification of the process at the NHanced foundry in 
North Carolina.  The initial NHanced run (Run 4) had poor oxide quality and large leakage currents. Therefore we were not able to fully qualify the SiSi process within the constraints of the R\&D program.

\section{Sensor and Test Structure Designs}
\begin{figure}[h]
\centering
\includegraphics[width=130 mm]{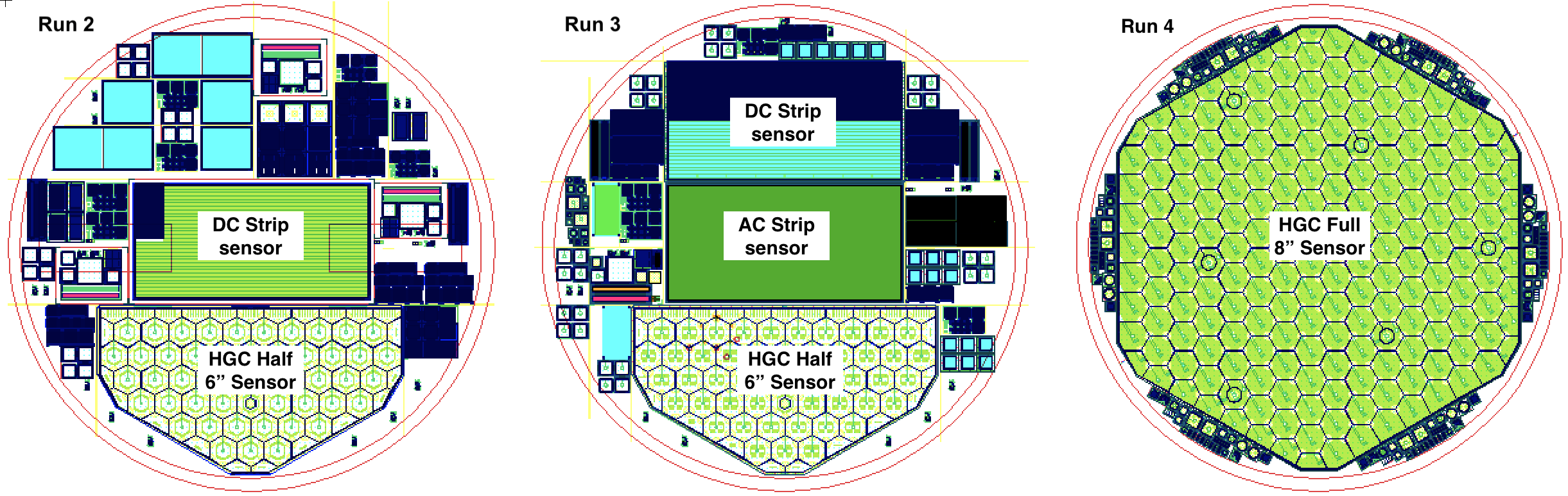}
\caption{Layouts for the three bonded wafer submissions.}
\label{fig:Runs}
\end{figure}

A 200 mm wafer provides ample space for both test structures and prototype designs. For Runs 2 and 3 we included a variety of designs from the High Energy Physics (HEP) and Basic Energy Sciences (BES) communities.  These included strip and pixel detectors for HEP and pixel sensors for 
x-ray imaging.  We also included a variety of test structures including pad diodes with guard ring variants, MOS test structures, and resistor and capacitor structures in the AC run. 
Figure \ref{fig:Runs} shows the overall layout of the wafers. Full size ($\approx 5 \times10 cm$) strip and macro-pixel sensors intended as prototypes for the CMS inner tracker pixel-strip (PS) module were included in runs 2 and 3. Run 4 was dedicated to a prototype 
full-wafer CMS High Granularity Calorimeter design.  

\section{Characterization Results}
\subsection{Depletion Voltage and Leakage Currents}
Here we consider only the devices from Runs 2-4, which incorporated wafer bonding technologies. The primary characterization tools were test structure diodes of $\approx 1\ cm^2$ . Samples of these diodes were VI and CV tested for several wafers in each run. Test structure diodes were IV tested to 600 or 800 V.  Test structures with currents exceeding $10\mu $A were considered to be in breakdown. We found substantial variations in breakdown voltage within a wafer, especially in Run 2. This made it difficult to establish statistically significant optimum values for the various process splits. 

\begin{figure}[h]
\centering
\includegraphics[width=120 mm]{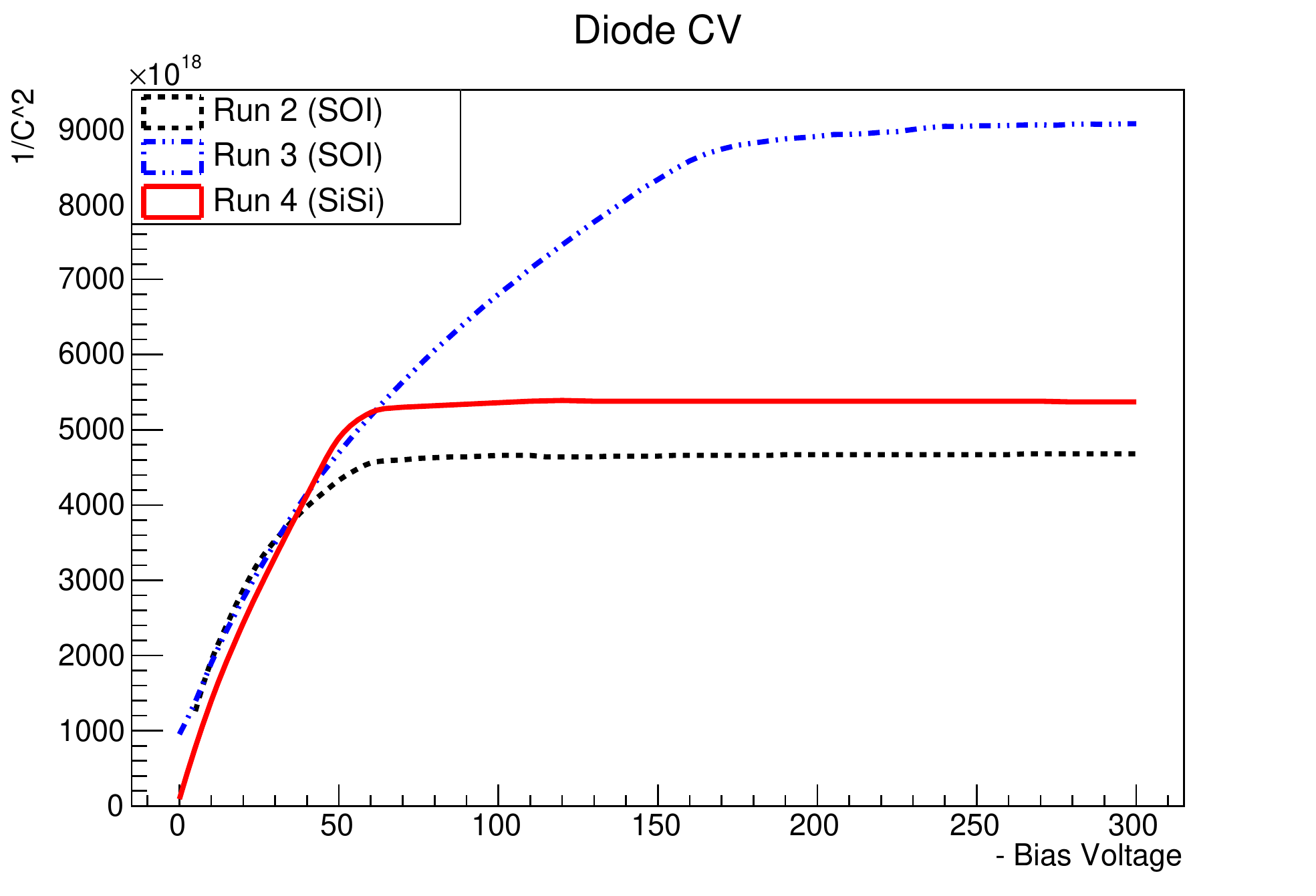}
\caption{$1/C^2$ vs $V_{bias}$ of Run 2, 3 and 4 test diodes. These are typical of diodes in each run.}
\label{fig:R3CV}
\end{figure}

\subsubsection{Run 2 Wafers}
Run 2 consisted of SOI wafers with the backside handle wafer intact.  The sensors on these wafers had to be biased from the topside contact since the backside implant is not accessible.
Diode CV tests showed a full depletion voltage of $60 \pm 10$ Volts (figure \ref{fig:R3CV}).  The measured active thickness is 200 microns. 
The calculated effective doping concentration is $2.0 \pm 0.3 \times 10^{12} /cm^3$. 

The leakage current at full depletion ranged from 0.5 to 0.75 $\mu A/cm^3$.  The range of breakdown voltage varied considerably from diode to diode within a wafer (figure \ref{fig:DIV}).  The devices processed with dry oxidation and p-stop implant dose of $5 \times 10^{11}$ had the best overall performance. In addition, some devices were observed to have hysteresis in the breakdown voltage, with the breakdown usually decreasing in subsequent iterations of the test. Devices with the single guard ring design generally had higher breakdown voltages by $\approx 100$ Volts with respect to the multi-guard designs tested.

\subsubsection{Run 3 Wafers}
Run 3 wafers were physically thinned and measured to have an active thickness of $250\mu m$.  These depleted at $170 \pm 15$ Volts, giving an effective doping concentration of $3.2 \pm 0.28 \times 10^{12} /cm^3$. The average leakage current at full depletion ranged from 0.18 to 0.3 $\mu A/cm^3$, considerably better than the Run 2 structures.
In general the breakdown voltage for these devices was higher and more uniform than the Run 2 sensors (figure \ref{fig:DIV}) and did not show hysteresis. We did not see a clear systematic 
difference between the various n-implant and p-stop process splits.

\subsubsection{Run 4 Wafers}
\begin{figure}[h!]
\centering
\includegraphics[width=100 mm, height=40 mm]{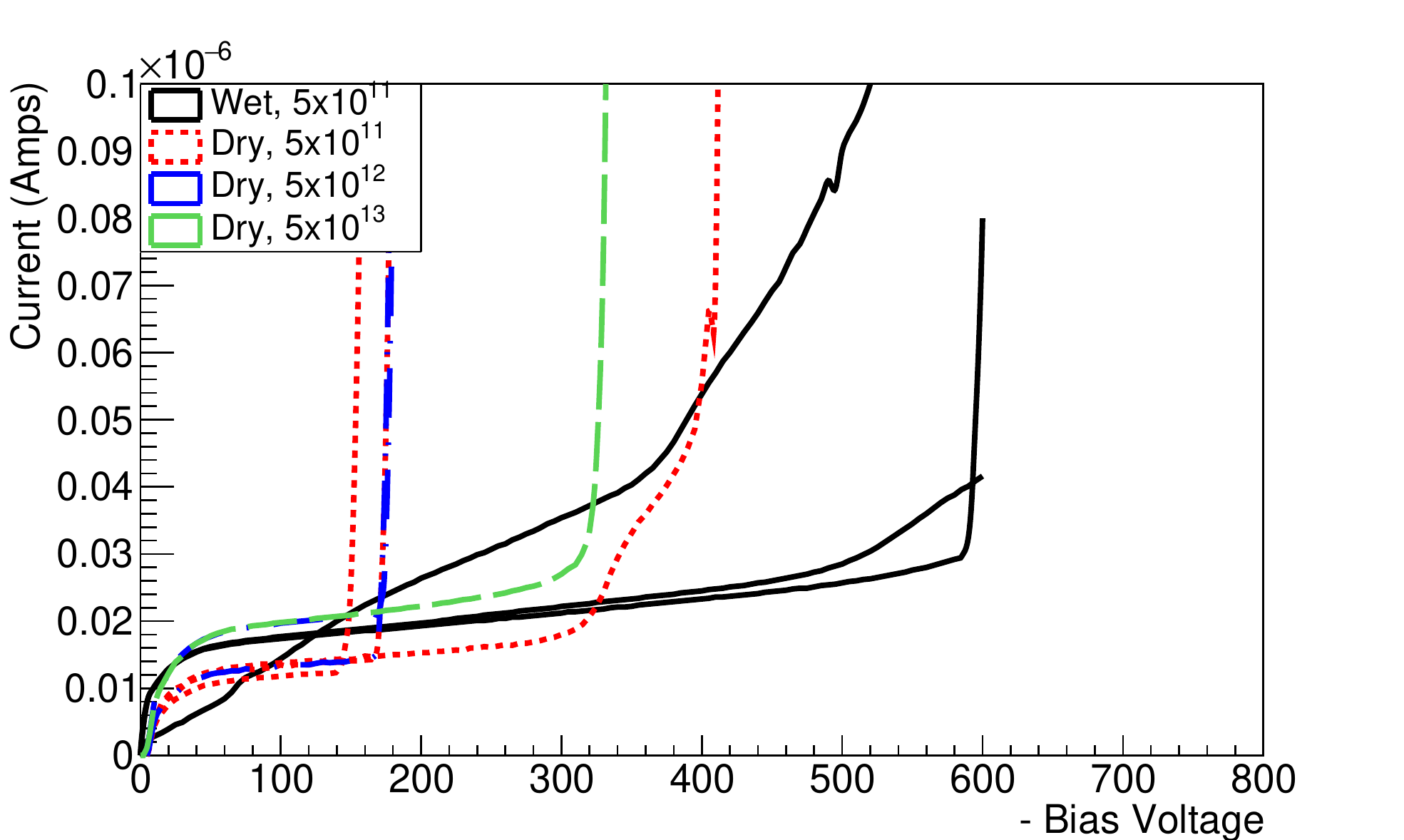}
\includegraphics[width=100 mm, height=40 mm]{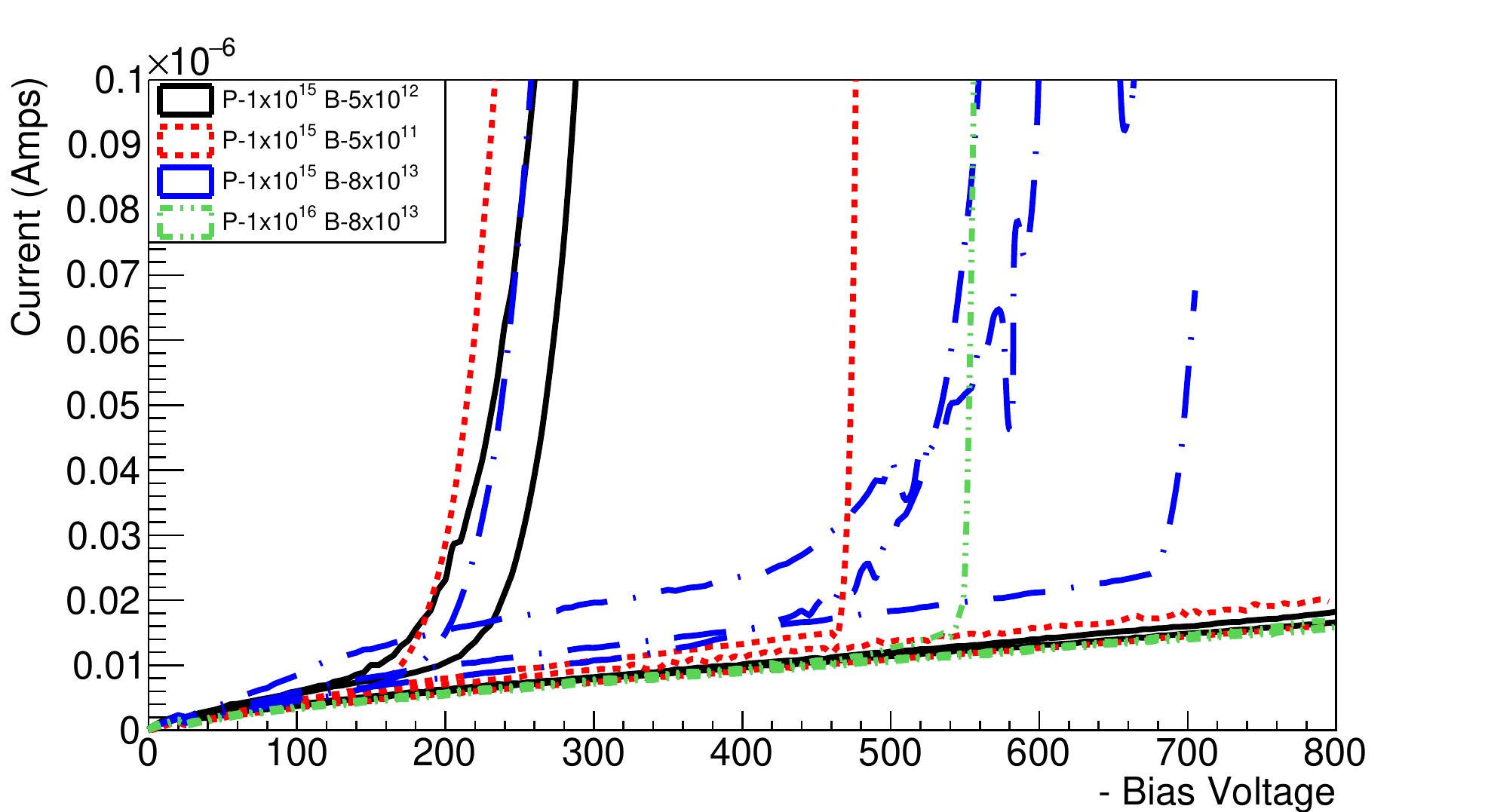}
\includegraphics[width=100 mm, height=40 mm]{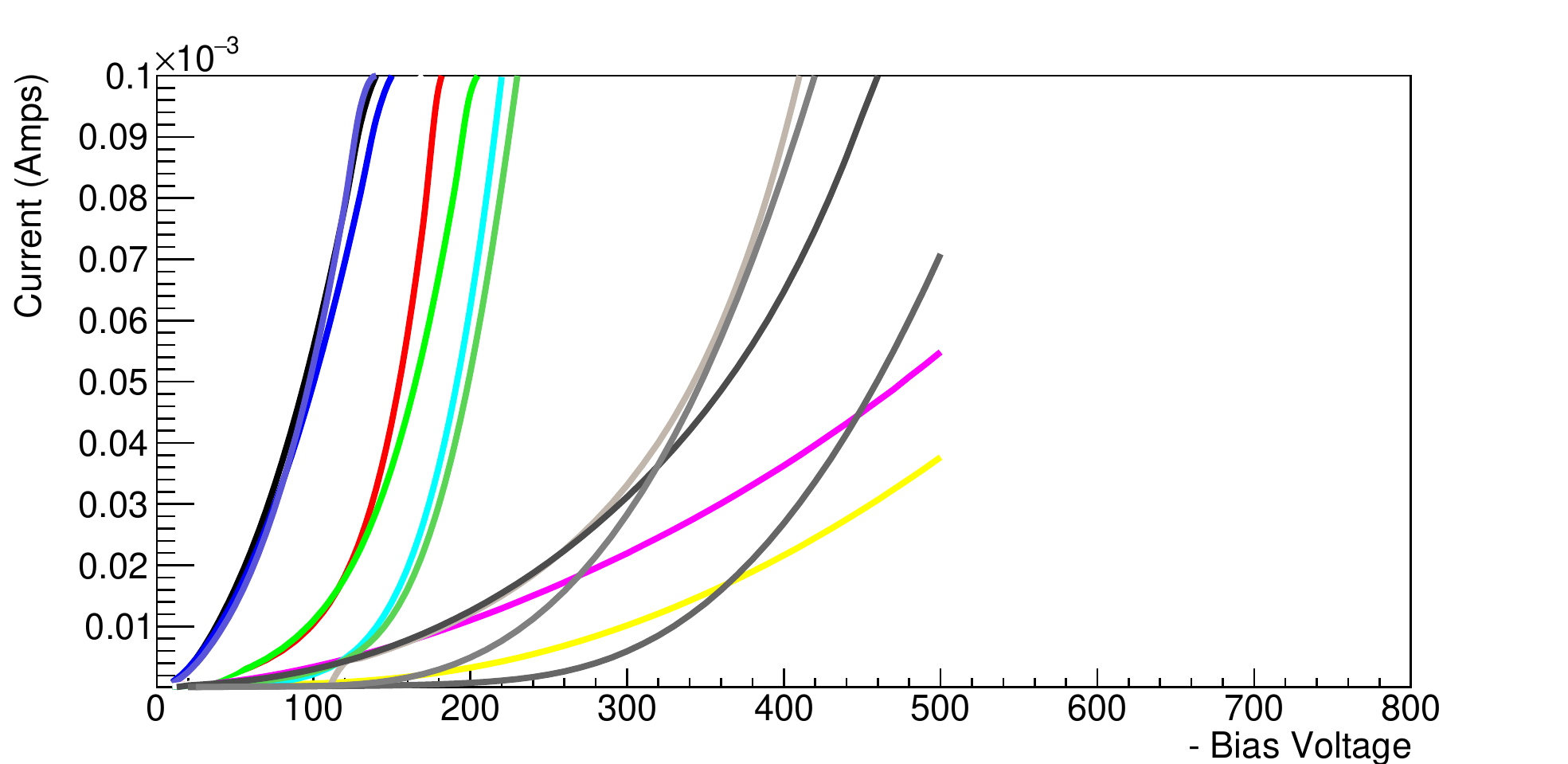}
\caption{One cm$^2$ test diode current vs voltage characteristics for diodes with a single guard ring for wafers from Runs 2 (SOI, top), 3 (SOI, middle) 
and 4 (SiSi, bottom). Note the compressed scale on the SiSi VI plot.}
\label{fig:DIV}
\end{figure}

Run 4 wafers were fabricated as a stack of low and high resistivity silicon. The high resistivity active region was measured to be 
200 microns thick. Depletion voltage for these sensors was  $60 \pm 10$ Volts (figure \ref{fig:R3CV}). The calculated doping concentration for 
the float zone layer is $2.0 \pm 0.3 \times 10^{12} /cm^3$. The leakage current is shown in figure \ref{fig:DIV}c. It is around $10 \mu A/cm^2$ at 20V above depletion, roughly
a factor of 10 higher than the SOI devices. The test diodes also show a rapid increase in current at 100-150 V, increasing to 0.1 mA at 
500 V.  The increase does not have the extremely sharp rise characteristic of avalanche breakdown.  This may be due to fields penetrating into 
the low resistivity wafer and the bond interface, drawing current from defects in these regions. A pre-bonding p+ implant in the float zone wafer could reduce 
this issue, but would also make the process more complex.

%\begin{figure}[h]
%\centering
%\includegraphics[width=73 mm]{R2_Diode_BD_LF.pdf}
%\includegraphics[width=73 mm]{R3_Diode_BD_LF.pdf}
%\caption{Breakdown voltage distribution for Run 2 and 3 test diodes. The %process splits are for various values of the p-stop implant dose.}
%\label{fig:R2BD}
%\end{figure}

\subsection{Surface and Interstrip/Pad Characteristics}
\begin{figure}[h]
\centering
\includegraphics[trim={0 5.5cm 0 4.5cm},clip,width=120 mm]{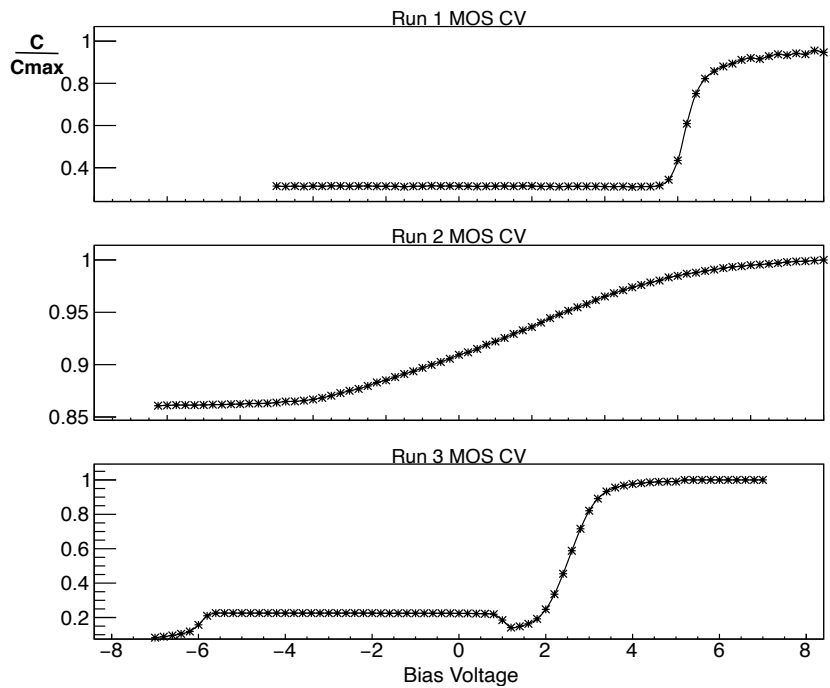}
\caption{MOS CV curves for Run 1,2 and 3 test structures. Capacitance values are scaled to the maximum value for each structure. Note the small range for the Run 2 structure. Run 4 structures did not yield a meaningful CV curve.}
\label{fig:MOSCV}
\end{figure}

Metal Oxide Semiconductor (MOS) test structure CV measurements are shown in figure \ref{fig:MOSCV}.  These measurements provide indications of the overall surface quality, including oxide and interface charges. The first three runs show distinct patterns. The first and third run show abrupt transitions between depletion and accumulation regions, indicating good interface quality. The small  slope in the Run 2 CV curve is indicative of problems in the bulk-oxide interface.  Run 1 had a large 6 Volt flatband voltage, indicating significant oxide charge density. This was improved for Run 3, with a 2.7 Volt flatband voltage corresponding to an oxide charge density of $1.2 \times 10^{11} /cm^2$. The MOS structures for the wafers in Run 4 did not show a clear distinction between the accumulation and depletion regions in the MOS 
test structures. Therefore the oxide charge could not be determined. This is true for both the Si-Si devices and float-zone controls. This is an indication of a poor silicon bulk to oxide surface interface.

%IV curves for Run 4 HGCAL pads were done by measuring the central (hexagonal) pad and grounding the six surrounding pads. Three of the four wafers tested showed a large increase in current, saturating the power supply, above a threshold of $\approx 300$V.  This appears to be due to poor inter-pad isolation causing charge to be collected on the six grounded outer pads from the surrounding floating pads. This was verified by biasing the surrounding pads and measuring the potential of the central pad. Initial interstrip resistance for the Run 4 test structures was measured to be low, a few times $10^4 \ \Omega$.  After full depletion the inter-pad resistance rose to $10^7$ to $10^9$ Ohms. All other runs had inter-diode resistance exceeding $10^8 \ \Omega$.

\subsection{AC Coupling Measurements}
Run 3 included polysilicon bias resistors and AC coupling capacitors. The polysilicon structures were implanted with a nominal phosphorous dose of $1 \times 10^{15} /cm^2$. The resistance of the polysilicon resistors ranged from 750 to 1000 Ohms per square. Serpentine resistors on the CMS PS-s sensors measured $207 \pm 4 \times 10^3 \ \Omega$, at the low end of the acceptable range for sensors of this type. AC coupling capacitors were designed with both oxide and nitride layers. For these devices the relevant quantitiy is capacitance per unit lenght. These capacitors were measured to have a capacitance of $80.9 \pm 0.2$ pF/cm. This is higher than the usual value of $\approx 20$ pF/cm, indicating a thin oxide/nitride dielectric.

\subsection{Irradiation Results}

\begin{figure}[h]
\centering
\includegraphics[width=73 mm]{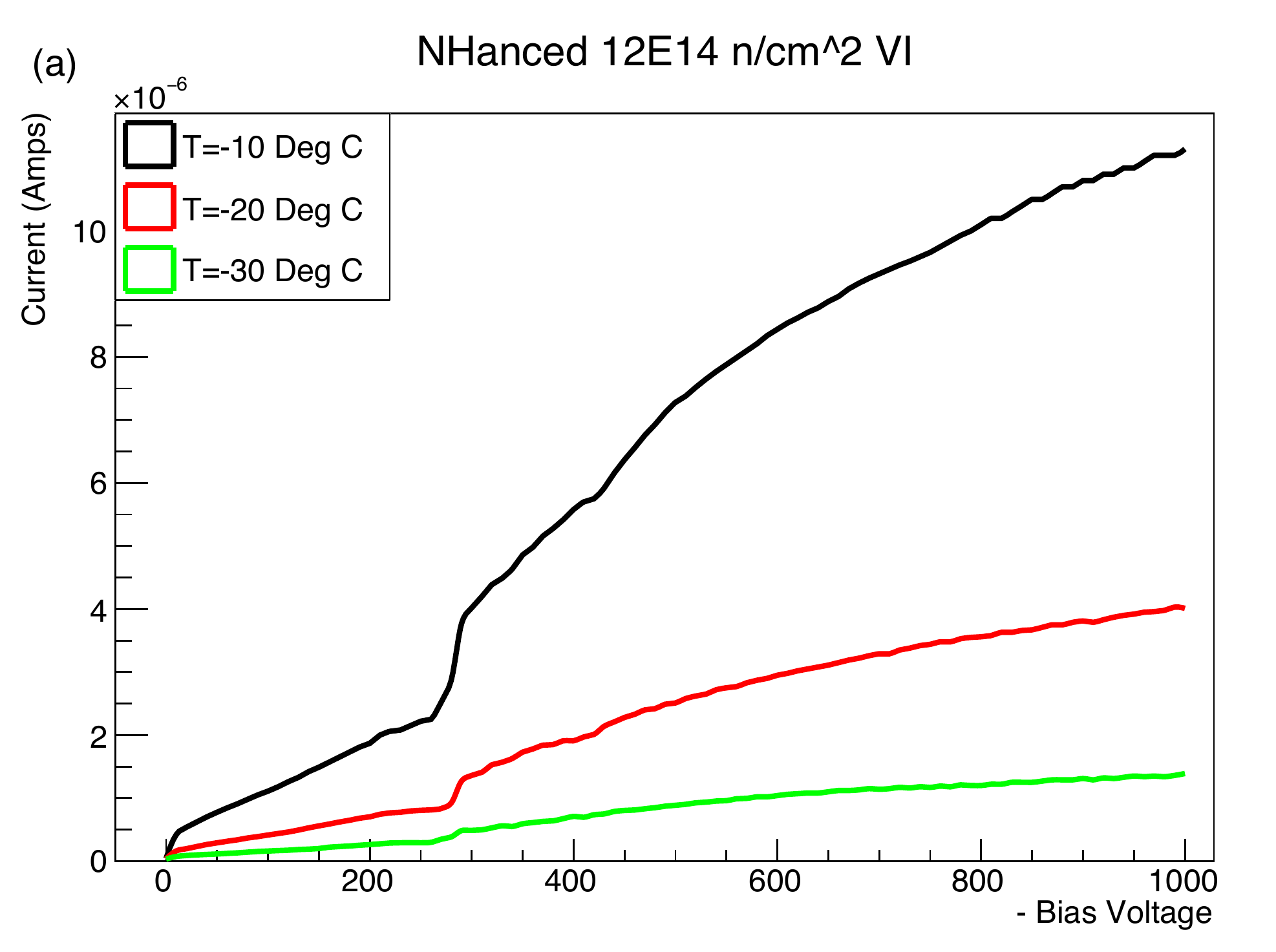}
\includegraphics[width=73 mm]{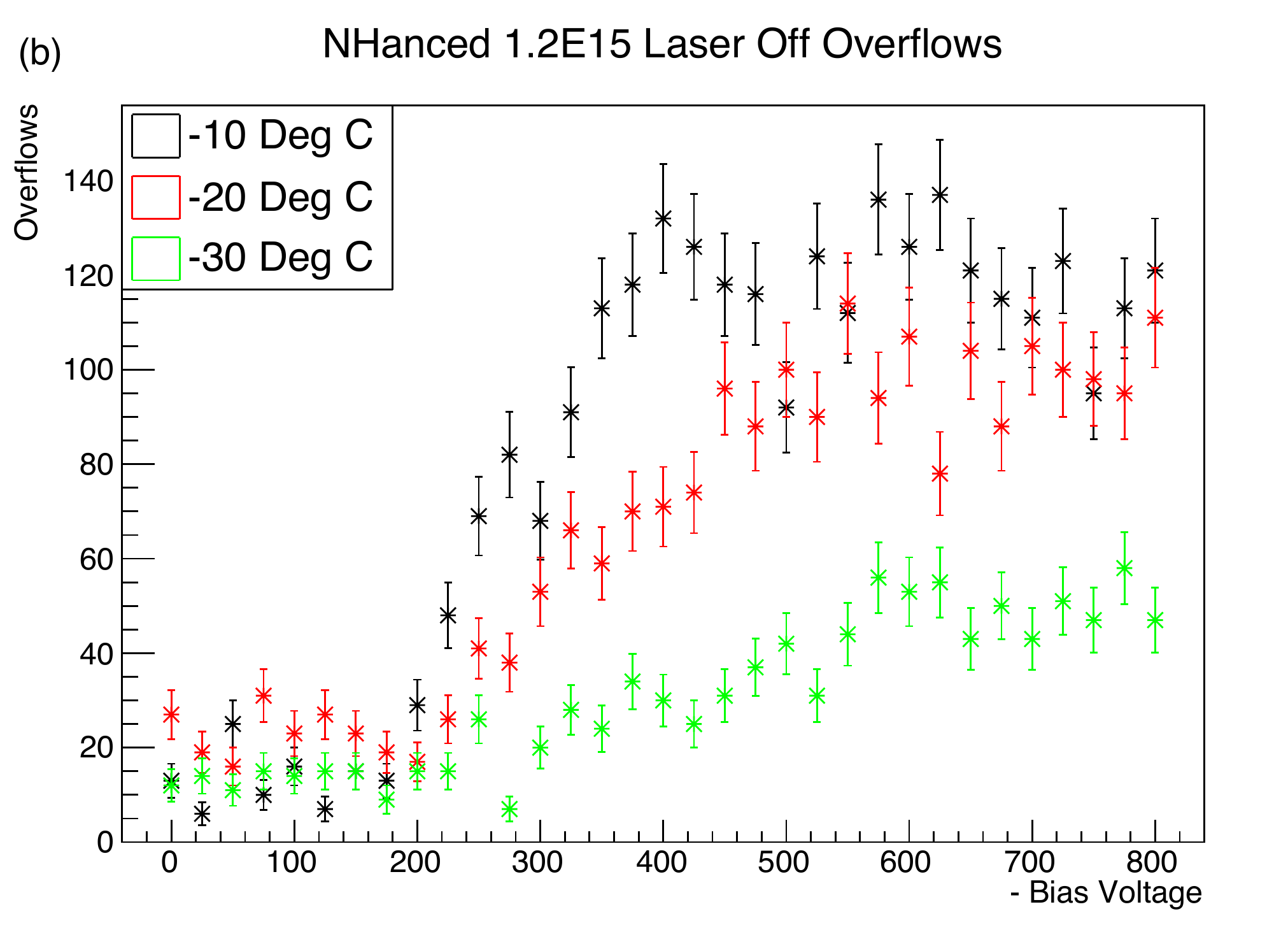}
\caption{ a) Current as a function of bias and temperature for an irradiated SOI HGCAL sensor. All curves show a break at about 300 V. b)Number of events with signal beyond 3 $\sigma$ of the pedestal for various sensor temperatures for the HGCAL sensor in 
a. There is a rapid increase in noise in the same region }
\label{fig:R3IVI}
\end{figure}

\begin{figure}[h!]
\centering
\includegraphics[width=100 mm]{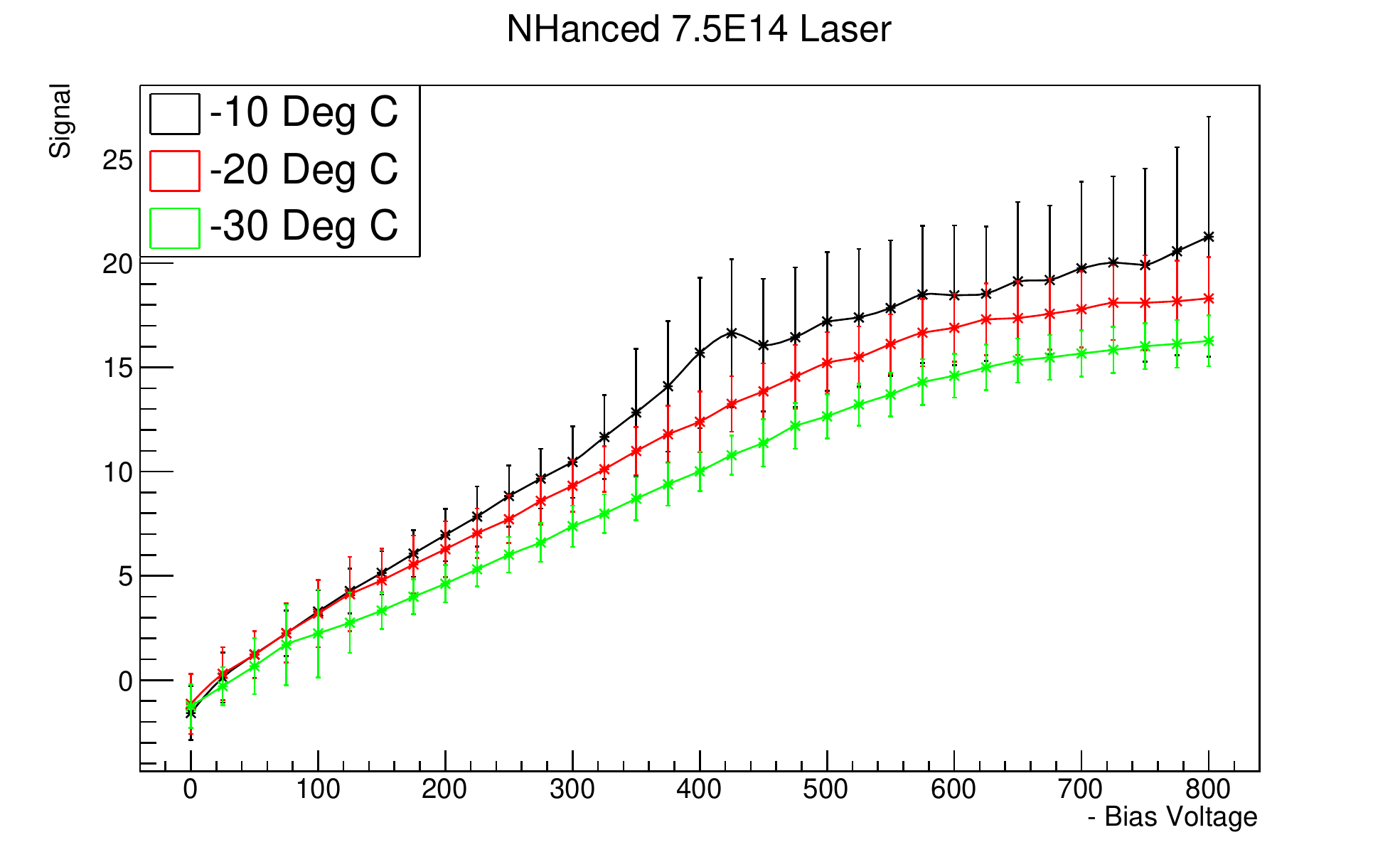}
\caption{Laser signal as a function of voltage for three values of detector temperature.}
\label{fig:R3Ilas}
\end{figure}

%\begin{figure}[ht]
%\centering
%\includegraphics[width=100 mm]{NHanced_12E14_LOff_Ofsum.pdf}
%\caption{Number of events with signal beyond 3 $\sigma$ of the pedestal for various sensor %temperatures for the HGCAL sensor in 
%figure ~\ref{fig:R3IVI}. There is a rapid increase in noise in the same region ($\approx %300$ V) as the current rise in the VI curves.}
%\label{fig:R3IOVF}
%\end{figure}

A Run 3 HGC half-sensor was included in an neutron irradiation run at the 
RINSC reactor in Rhode Island.  This run had an estimated 1 MeV neutron equivalent 
fluence of $1.2 \times 10^{15}\ n/cm^2$. In addition to the usual VI tests these devices were tested for charge collection with a 1064 nm laser and transimpedance amplifier. An 
200 mm semi-automatic probe station equipped with a thermal chuck was used. A seven pin probe card provided contact to the central pad of the hexagonal array while maintaining ground potential in the six surrounding pads.  Non-Gaussian noise is measured by counting events beyond 3 $\sigma$ of the pedestal. This is an indication of possible breakdown. 

Voltage-current curves for the irradiated Run 3 sensor are shown in Figure \ref{fig:R3IVI}a. The current ratios are consistent with the standard temperature dependence of leakage current. 
Figure \ref{fig:R3IVI}b shows the number of non-Gaussian noise events in runs with the 
laser off. There is a increase in noise in the region 
between 200 and 300V. 
We note that this corresponds to a break in the leakage current
VI curves in that region. We conclude that there is a possible 
onset of breakdown in this region.

Laser test charge collection results for the Run 3 sensor are shown in figure \ref{fig:R3Ilas}. We believe that the variation in charge collected with temperature is due to the temperature variation of the absorption coefficient for infrared light at this wavelength\cite{G1}\cite{Maral}. The charge collection appears to plateau at 600-700 Volts. 
The calculated in depletion voltage is $\approx 750$ V for this fluence, including an estimate of the annealing in the reactor during exposure.

\section{Conclusions}
We have demonstrated the fabrication of 8" sensors with thin active regions using both SOI and SiSi bonded wafers. SOI bonded wafers with the handle wafers removed (Run 3), provided the best results, with acceptable leakage currents and  breakdown voltage values and  consistency. This run also demonstrated AC coupling resistors and capacitors with acceptable characteristics. A sample of Run 3 sensors were irradiated to $1.2 \times 10^{15}\ n/cm^2$. These showed the expected leakage current and depletion characteristics. Evidence was found for a non-Gaussian component of the noise above 300 Volts. We were unable to qualify the Si-Si process in these studies due to changes in fab site and processing. The SOI process worked well and shows promise for future development of sensors with thin active layers.

%There was not enough distinction in results among wafer splits tested to make definitive statements regarding the optimal oxydation process, p-stop dose, and n-implant dose. Our best results in the first three runs were obtained with dry oxide, and a p-stop dose of $5 \times 10^{13}$. 

%After irradiation of the Run 3 wafers there appears to be a noise source, presumably due to the onset of breakdown, which takes affect around a bias voltage of 300 volts.  The magnitude of this effect drops with temperature.

\section{Acknowledgments}
This manuscript has been authored by Fermi Research Alliance, LLC under Contract No. DE-AC02-07CH11359 with the U.S. Department of Energy, Office of Science, Office of High Energy Physics. Partial funding for this work was provided by the DOE Small Business Innovative Research (SBIR) program

We would like to thank the summer students (P. Camporeale, J. Everts, R. Halenza, and A. Lampis) who participated in making  measurements of the sensors and test structures.

%\clearpage

%\section*{References}

\bibliographystyle{plain}
\bibliography{8in}

\end{document}